\documentstyle[12pt]{article}

\begin{document}

\title{Reply to comment on \\ ``A simple explanation of the non-appearance of
physical gluons and quarks"}
\author{Johan Hansson \\ Department of Physics \\ Lule{\aa} University of
Technology \\ SE-971 87 Lule\aa, Sweden
\vspace{0.5cm}\\
{PACS numbers: 12.38.Aw, 03.70.+k, 11.15.-q} }

\date{}

\maketitle

\begin{abstract}
This is the reply to a comment by Andreas Aste on a previous
article of mine in Can.J.Phys. The counter-arguments used by Aste
utilize a mathematical limit without physical meaning. We still
contend that in QCD, the particles ``gluons'' and ``quarks'' are
merely artifacts of an approximation method (the perturbative
expansion) and are simply absent in the exact theory.
\end{abstract}
The critique of my previous article \cite{Hansson} by Aste
\cite{Aste} centers on scattering theory (S-matrix) and plane wave
solutions to the evolution equation
\begin{eqnarray}
(\delta_{ab} \partial^{\mu} + g_s f_{abc} A_c^{\mu})
(\partial_{\mu} A_{\nu}^b - \partial_{\nu} A_{\mu}^b + g_s f_{bde}
A_{\mu}^d A_{\nu}^e) = 0,
\end{eqnarray}
for color fields without quark sources.

The definition of the scattering matrix is that it transforms
\textit{free} states (harmonic oscillator mode excitations =
particles) at $t=- \infty$ into \textit{free} states at $t=+
\infty$. In other words one has to \textit{pre-suppose} the
existence of asymptotic states, i.e. plane waves (which of course
can be quantized in the usual manner). Also the interaction
Hamiltonian, $H_I$, is (iteratively) made up of such free field
operators.

The eight different uncoupled gluons (eq.(3) in Aste's paper
\cite{Aste}) resulting from plane wave solutions to eq.(1) above,
is solely a result of an unphysical mathematical limit. These
asymptotic states are never physically realized, as it either
demands that the coupling is identically zero ($g_s \equiv 0$)
leading to a non-dynamical theory, or that $Q^2 \rightarrow
\infty$ where perturbative quantization is feasible through
asymptotic freedom. The latter, however, does not imply physical
particles observable in the lab. (This should be contrasted to the
case in QED, where asymptotic, and physical, states are reached as
$Q^2 \rightarrow 0$.) The sterile character of the solutions can
also be seen directly from the linear plane wave form of eq.(3) in
\cite{Aste}, as it is well known that linear waves do not
interact. A more technical way of saying the same thing is that
this special solution demands the absence of local gauge
invariance (``the color of the gluon field is held constant
throughout the space" \cite{Aste}) which implies non-interacting
fields.

However, why introduce these unphysical states at the start and
then, unsuccessfully, try to get rid of them in the end (to
explain ``confinement"), when they do not appear in the full
non-perturbative theory?

Hence, the counter-arguments in \cite{Aste} are not physical, and
we still contend that the nonappearance of \textit{physical}
gluons and quarks is a direct consequence of QCD, as shown in
\cite{Hansson}.

\end{document}